# SHoM: A Mental-Synthesis Trust Management Model for Mitigating Botnet-Driven DDoS Attacks in the Internet of Things

Masoud Hayeri Khyavi
Qazvin Azad University
Qazvin, IRAN
m.hayery@itrc.ac.ir

## Abstract

The advantages of IoT in strengthening commercial, industrial, and social ecosystems have led to its widespread expansion. Nevertheless, because endpoint devices have limited computation, storage, and communication capabilities, the IoT infrastructure is vulnerable to several cyber threats. As a result, DDoS attacks pose a significant risk to the security of IoT devices. By exploiting these weaknesses, attackers can quickly utilize IoT devices as part of botnets to execute DDoS attacks. The most critical development is the construction of more armies of robots from IoT devices. We offer a Model for dealing with DDOS attacks on botnets in the Internet of Things via trust management. In this Model, an attempt has been made to consider all aspects of security related to trust factors in designing a reliable and flexible model against DDoS attacks on the Internet of Things. In initial studies, approximately 40-50 security models related to the subject were reviewed using review articles.

## 1. Introduction

Because one of the unresolved problems in the Internet of Things security is the trust issue between various elements in the Internet of Things, choice management in this context is crucial. On the other hand, it is critical to examine DDoS attacks on the Internet of Things because they have impacted all of the Internet of Things' layers. DDoS attacks are centred on botnets. The IoT ecosystem is not only highly vulnerable to DDoS attacks, but it may also be utilized to launch DDoS attacks against other targets.

Concerns about the value of security and trust have been raised by several researchers. Different ideas are addressed for security concerns, security procedures, security technologies, and security services related to the trust issue. One of the fundamental prerequisites for security and security methods for developing IoT networks is trust management, which is essential for enhancing security and user privacy. Additionally, trust management is a simple solution that offers protection for IoT devices. For these devices, trust management methods based on subjective logic have also been developed. IEEE 802.15 evaluates these algorithms.

Trust management aims to secure access control and prevent, detect, and neutralize malicious nodes in the Internet of Things. Authentication flaws can be covered via trust management. We have to deal with the trust management system and trust management strategies when it comes to managing trust in the Internet of Things. An IoT system might be a collection of guidelines that restructure how IoT applications are used. Because it offers security for all layers and networks, trust management models play a significant role in the Internet of Things in protecting data and devices from attacks.

The most important areas for future study are "malicious node detection," and the minimum importance is related to "trust-based access control." However, the Internet of Things architecture was designed with the following security levels:

- End-to-end security (although the equipment need not have mutual trust at this level, there is trust management.)
- Edge-created security services
- Distributed security model

Additionally, the threat model is a technique for determining, quantifying, and researching the security threats connected to a system, including the Internet of Things. Using the threat model as a good beginning point will help comprehend the risk related to these systems and learn ways to lower these risks. There are several threat model frameworks available. This paper is an abstract of the author's research phase in preparing the Ph.D. proposal. In the following articles, the formulation of the Model, the results of the implementation of this plan in the laboratory, and a comparison with other sample plans will be presented. The contribution of this Model can be found in the examination and combination of different technologies. Ruled modelling is the crucial link in the operation mechanism of cyber situational awareness. In the proposed Model (SHoM), all aspects have been considered, and a model has been presented that can be used as a basic model in future research and the use of emerging technologies in the field of management. Trust, counter botnets to be used. Also, this Model can be examined and improved by experts in the field of Internet of Things security.

## 2. Theoretical foundations

***A) Trust management components***

In Figure 1, five components are seen. We now go over each of the parts as mentioned above in detail:

### I. Trust composition

It refers to the components considered when calculating trust and consists of two primary modules: Quality of Service (QoS) trust and social trust.

- An IoT entity is expected to deliver higher quality in its operations, referred to as QoS. To quantify the value of trust, QoS trust employs several trust metrics, including competence, reliability, task fulfilment, and cooperation.
- Social relationships between IoT entity owners are referred to as social trust. The Internet of Things is evaluated based on social relationship trust to determine its reliability. Additionally, social trust measures trust values using trust features.

## II. Trust formation

This refers to using a single trust feature for multiple trust features for the trust computation. Additionally, these elements primarily address how much weight is assigned to service quality and social trust features.

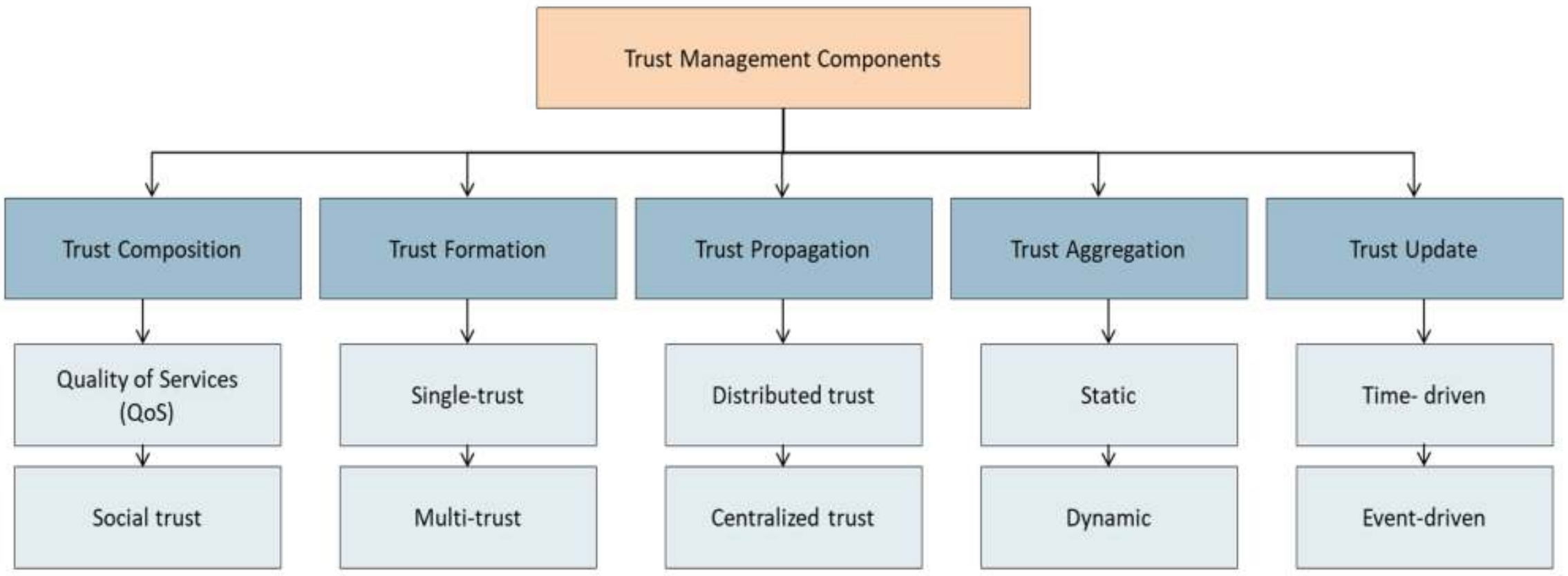

**Figure 1. The trust management model components [1]**

## III. Trust Propagation

Trust distribution information to other entities is referred to as trust distribution. Two preliminary designs fall under this category of distribution:

1. The term "distributed trust" describes IoT that autonomously shares its trust and observations with other IoT entities it interacts with or directly meets without needing a centralised organisation.
2. A physical cloud implemented by IoT devices or a virtual trust service is an example of centralised trust, which implies the existence of centralised organisations.

## IV. Trust aggregation

This component refers to the most appropriate method of collecting trust information, after which the result is directly or indirectly evaluated by the entity itself. This component collects data using the weights taken into account, which may be static or dynamic, with the static case computed under entity features. The initial trust between the two communication parties is based on each party's reliability. When allocating weights to each element, the trust manager must use contextual information to make optimal dynamic trust judgments. Different models of trust aggregation, such as belief theory, fuzzy logic, Bayesian inference, weighted sum, and regression analysis, are discussed in the relevant literature.

### V. Trust update

The time for updating the trust values is decided by this component. Updating trust information occurs periodically (time-driven) by applying a trust aggregation or after a transaction or event affects QoS (event-driven). Nineteen models have been investigated in related models and policies or trust management; however, we study each Model and analyze its strengths and weaknesses to theoretically compare our suggested Model with them by detecting them.

## 3. Cybersecurity awareness models

*Awareness of the cybersecurity situation*

The following is mentioned in [2]'s definition of network position awareness in cyberspace:

The term "Cyberspace Situation Awareness" (CSA) refers to the extensive network environment where decision-making and Action are the main objectives. This theme's basic structure is composed of three levels and three layers:

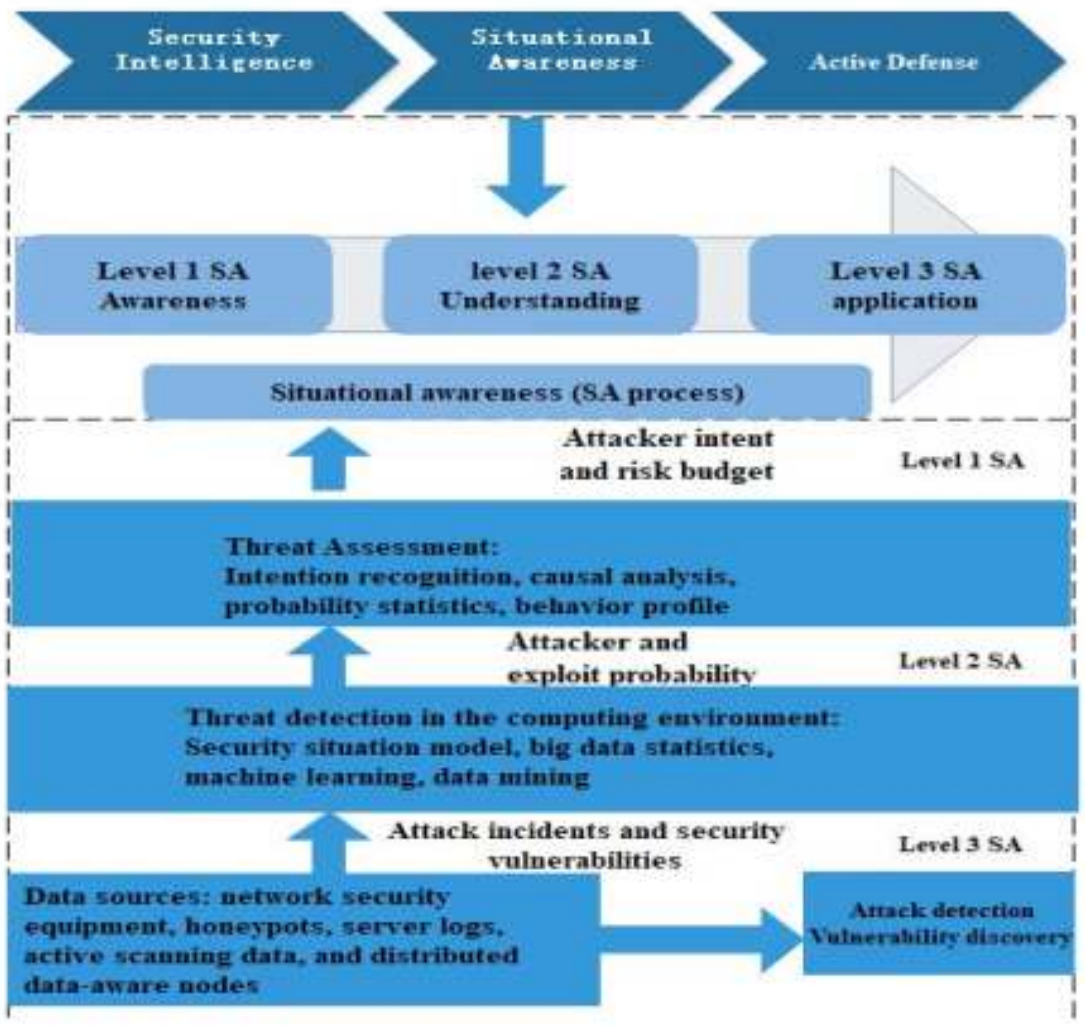

**Figure 2. The three-layer architecture of situational awareness**

Figures 3 to 6 show, accordingly, the several models that have been discussed and contrasted in this article's discussion of network situation awareness:

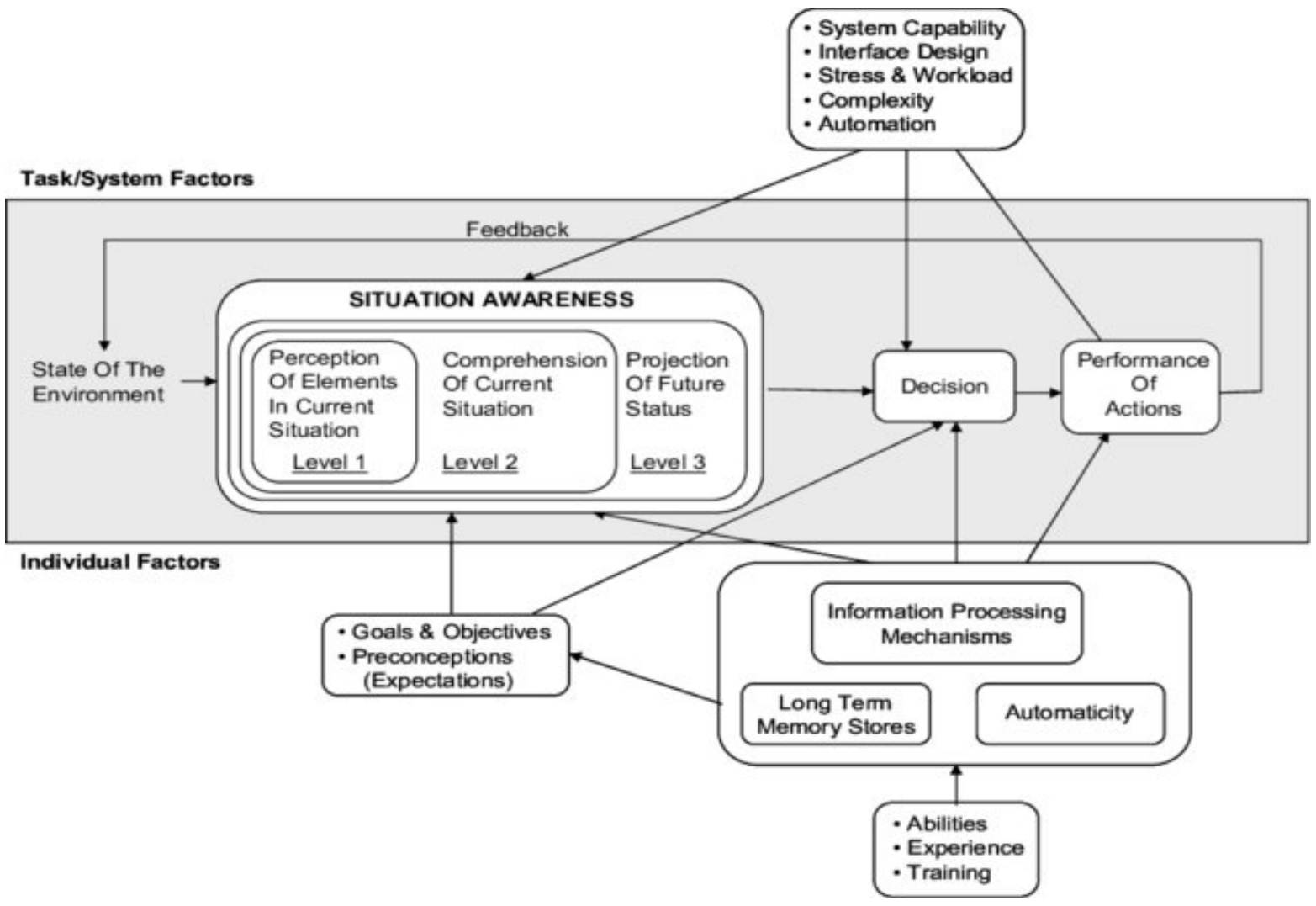

**Figure 3. Endsley model**

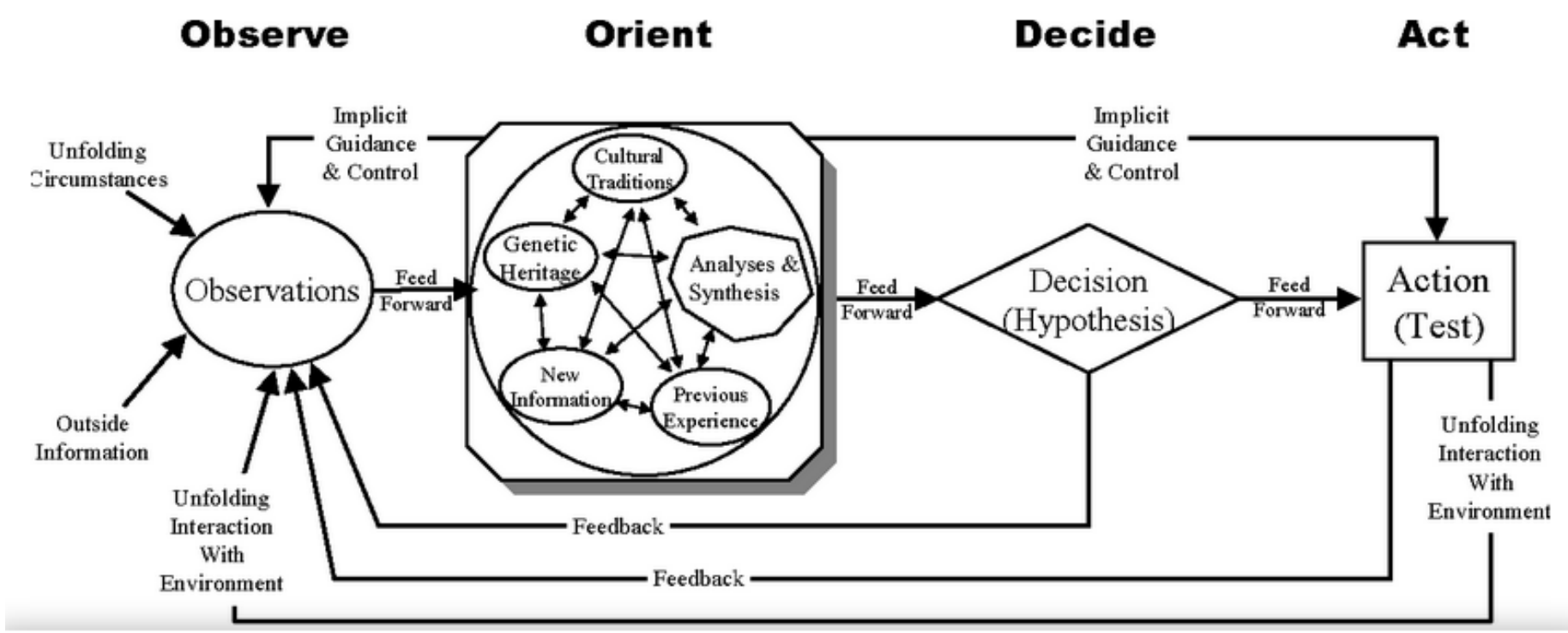

**Figure 4. Observe Orient Decide Act (OODA) model**

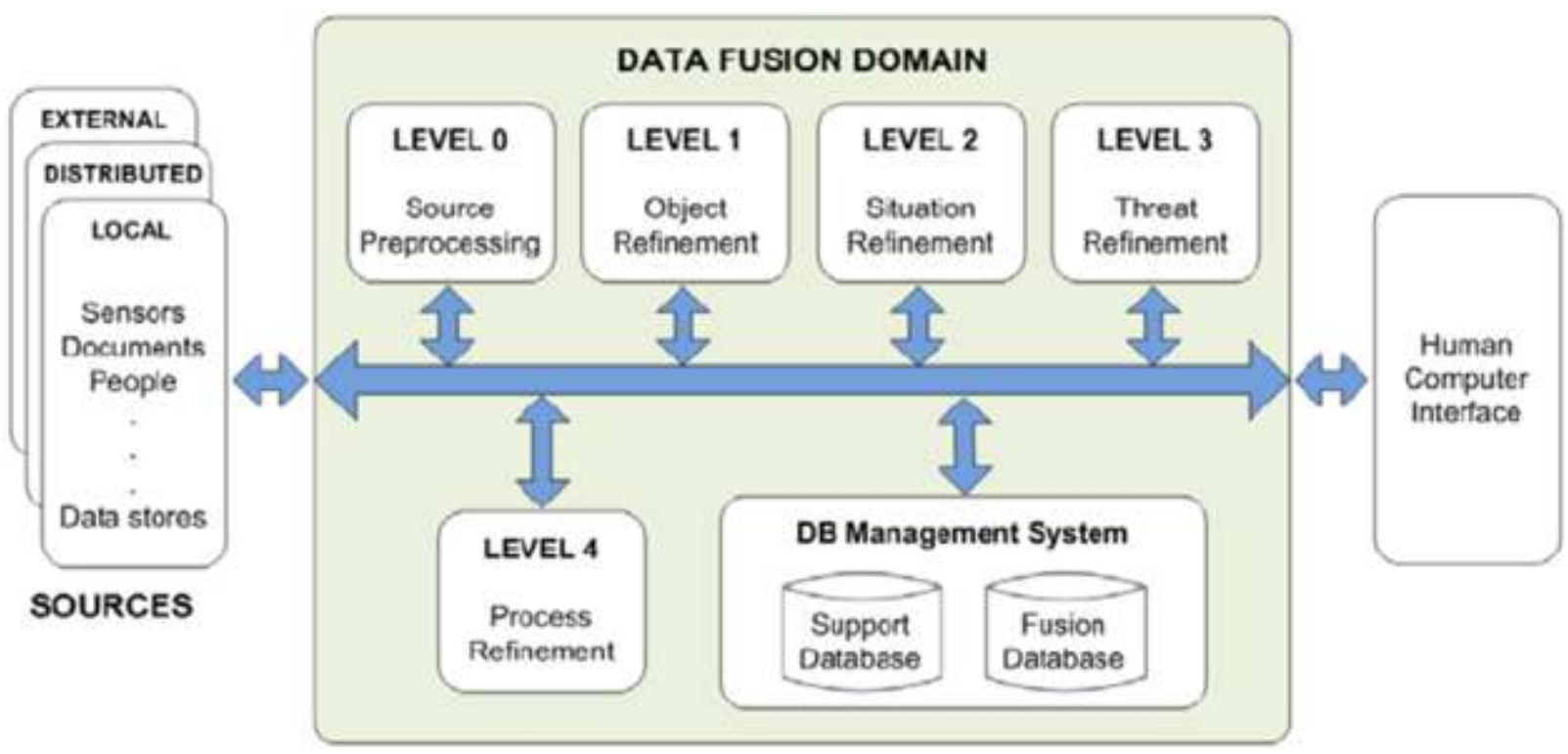

**Figure 5. Joint Directors of Laboratories (JDL) Model**

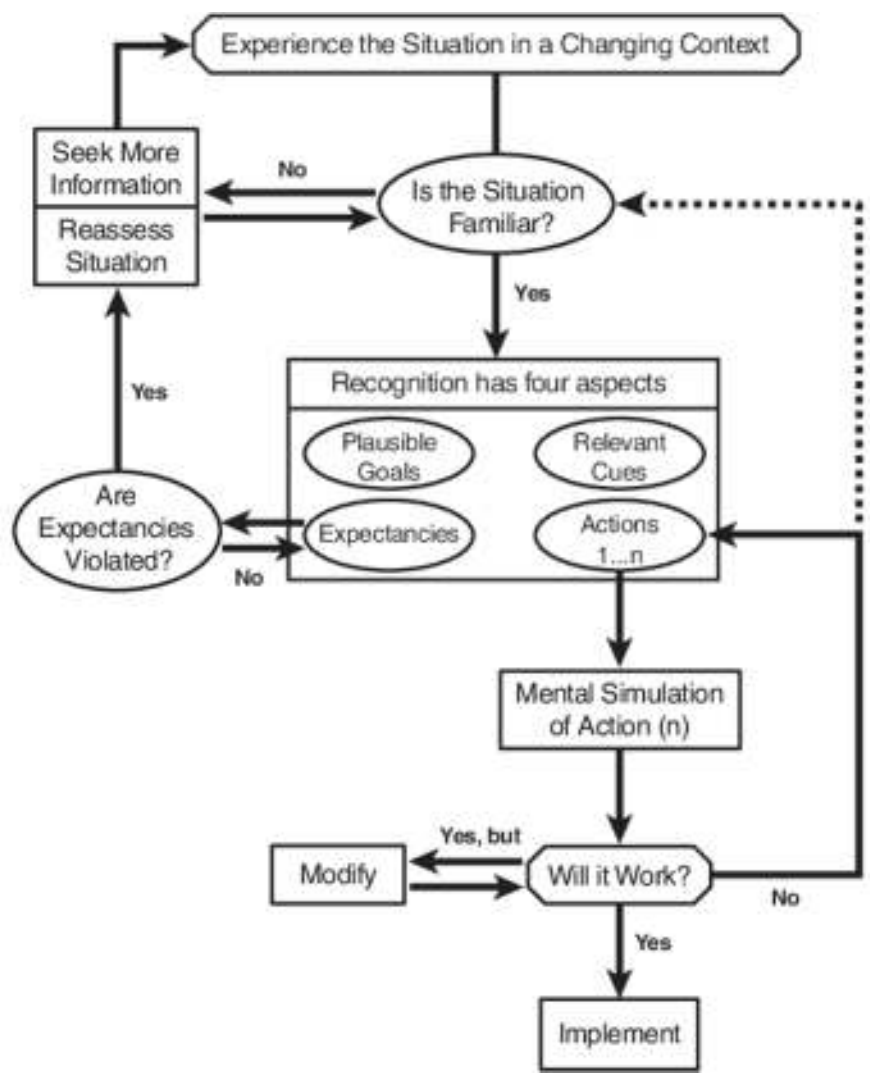

**Figure 6. The Recognition Primed Decision (RPD) model**

A deep learning network-based model is presented in [2].

## 4. Related Work

Given the extensive scope of the reviewed literature, this section synthesizes key findings and issues from selected critical studies:

1. Lack of integration of dynamic trust management: Most models consider trust as an independent component, not as the core of the security system [3, 4].
2. Inflexibility in dynamic environments: Current solutions lack self-organizing mechanisms to adapt to evolving attacks and changing network conditions [5, 6].
3. One-dimensional approaches to DDoS: Most models focus on one attack layer (e.g., volume) and neglect other dimensions (protocol/application) [6, 7].
4. Reliance on centralized architecture: Solutions based on a central controller (e.g., SDN) are vulnerable to single-point attacks [6, 8].
5. Failure to integrate human-technical dimensions: Current models ignore perceptual concepts (e.g., contextual analysis of device behaviour) [4, 9].
6. Scalability challenges in IoT: Computationally intensive algorithms (e.g. deep learning) are incompatible with the resource constraints of IoT devices [6, 10].
7. High false positive rate: Anomaly-based systems have low accuracy in heterogeneous IoT environments [6, 11].
8. Neglecting the special nature of IoT traffic (MTC): Traditional solutions overlook the differences between machine-to-machine (MTC) and human-to-machine (HTC) traffic [12].

## 5. Research Methodology

It has been noted in [13] that a systematic method is recommended for cybersecurity-related processing of data and that situation awareness of network security often encompasses multiple different phases. Although none of these approaches can offer an easily detectable architecture from the point of view of the data processing stage, the same source notes that there are two basic ways for logical partitioning in this context:

1. Hierarchical method of engineering
2. Conceptual hierarchy method

The authors in [13] have taken a systematic engineering approach from the perspective of the data value chain, which they assert has been widely adopted by the industry. Figure 7 displays this Model.

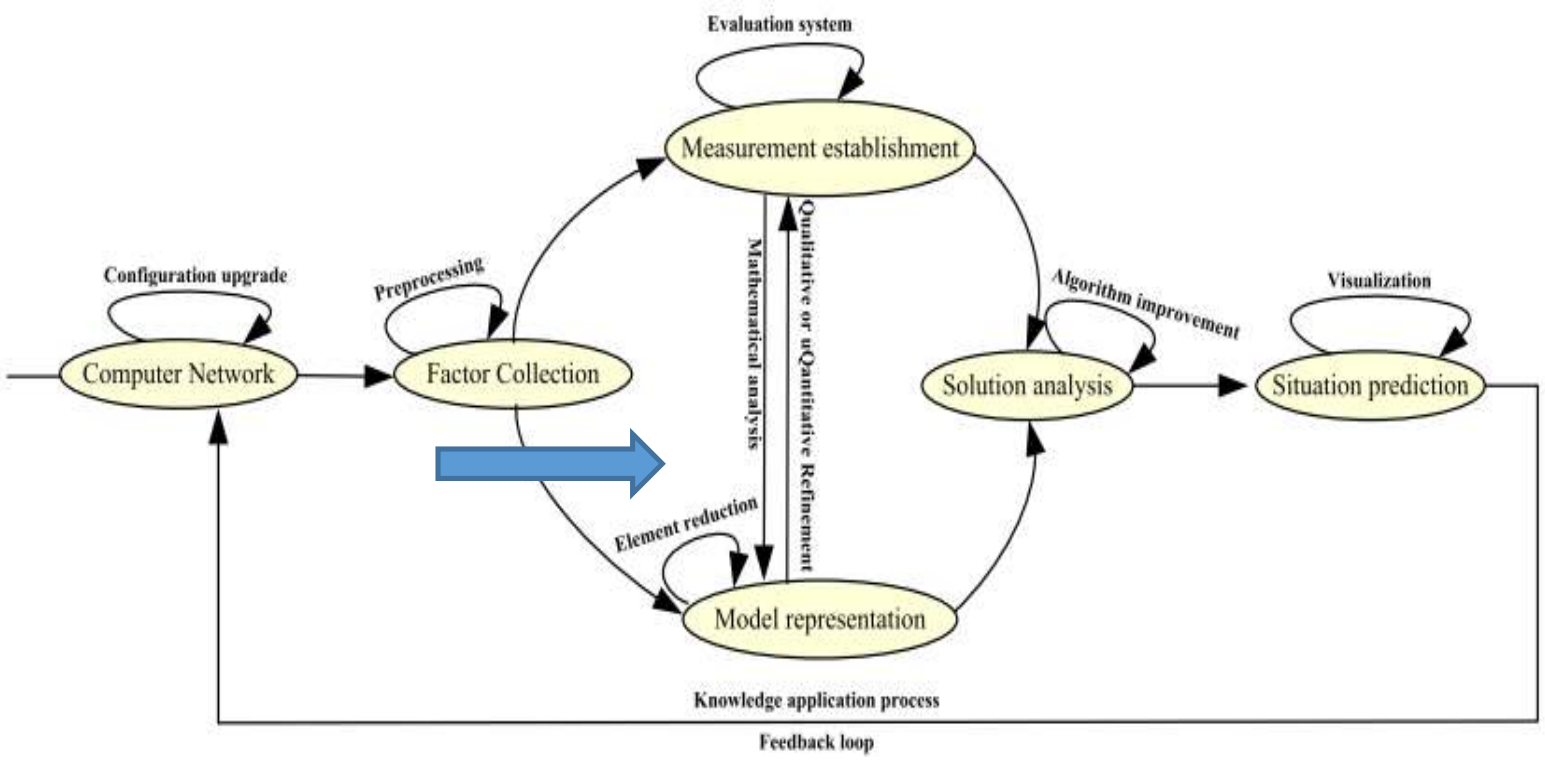

**Figure 7. The mechanism of network security situational awareness operation**

Considering the above figure (Fig.7), the following findings have been obtained [13]:

**1. Factor collection**: collecting network configuration data, activity data from logs, and vulnerability data that may be accessed using a scanner, sensor, or specialized tool.

**2. Data Preprocessing**: It is the process of regularising the original data before modelling, analyzing, and exploiting the data.

**3. Model representation**: It has two tasks: conversion/reduction of elements (more effective) and formal representation (accurate feature-based summarisation, finding the elements' relationship and order relationship.

**4. Metric establishment**: It is the process of modifying/fine-tuning the number of objects (tangible objects) that represent the element objects before analyzing the solution, mainly including quantitative classification (allocation of feature values numerically) and a measurement index system to determine two activities.

**5. Solution analysis**: It is an algorithmic process based on the first three steps discussed above, which primarily entails determining, verifying the accuracy, and enhancing the algorithm tasks.

**6. Situation Prediction**: This prediction is a thorough examination and decision-making process based on the analysis's findings. It primarily entails two tasks: visualizing the results and choosing after applying the information. After completing these two processes, a perceptual loop is closed by using a feedback loop to the existing network to improve cybersecurity (vulnerability repair and configuration updates). In Figure 7, the stage of mathematical analysis is shown with a dark arrow (from measuring to presenting the Model), and it is defined in two phases:

**Phase one: element acquisition**: This stage's purpose is to efficiently collect critical data at each level of cyber situation awareness. This phase, in general, refers to containing all cybersecurity-related components, which are divided into three sections:

- Data generation
- Data acquisition
- Data preprocessing

These data are classified as follows:

**A. Static data**

**Table 1. Static data**

| Host | Network | IDS |
|---|---|---|
| ○ identity<br>○ service<br>○ assets<br>○ operating system<br>○ hardware<br>○ configuration<br>○ Access permissions | ○ Topology<br>○ protocol<br>○ firewall<br>○ configuration | ○ Basic data<br>○ Dependent/relevant knowledge<br>○ Warning data |
| **Procedure:**<br>Manual and automatic | **Procedure:**<br>automatic | **Procedure**:<br>Manual and automatic |
| **Sample tool:**<br>HostScan<br>SuperScan | **Sample tool:**<br>MyLanViewer<br>NetX<br>IP Scanner | **Sample tool:**<br>Snort<br>Nmap<br>Nikto |

**Table 2. Dynamic data**

| Received result | Attack | Vulnerability | Behavior | Activity |
|---|---|---|---|---|
| o Description<br>o Action | ○ Resource<br>○ Method | o Basic data<br>o Host data<br>o Attack method<br>o Damage<br>o Repair method | Source<br>destination<br>protocol<br>Data<br>Encryption algorithm | Source<br>Destination<br>Action |
| Procedure: Manual | Procedure: Manual and automatic | Procedure: automatic | Procedure: automatic | Procedure: Manual |
| **Sample tool** ---- | **Sample tool** Snort | **Sample tool** ISS Scanner Whisker Nessus | **Sample tool** MyEventViewer LogFusion | **Sample tool** ---- |

**Phase 2: Model presentation:**

A crucial component of cyber situation awareness activities is rule-based modelling. The results of the subsequent perceptual analysis are directly impacted by this phase. Three categories [13] are used to categorize the models.

***1) Mathematical Model:***

The analysis of cybersecurity situation awareness uses a mathematical model. The main idea of using mathematical language or mathematical symbols is to summarise or approximate the dependencies related to security or quantify the dependencies of computer network systems. The mathematical formulation of the interactions between the variables of the cybersecurity system serves as the context-specific mathematical Model in this context. As a result, a quantitative analysis is required.

***2) Random/stochastic Model***

A non-deterministic model is the stochastic analysis model. Its main feature is that the external variables in the Model will change with certain conditions that will have a high degree of fit with the occurrence of cybersecurity-related behaviours. A stochastic model may be used to demonstrate the logical connection between random behaviour and the activities of various system components in cybersecurity situation awareness.

***3) Biologically inspired Model***

The biological heuristic calculation approach is a model that draws inspiration from natural occurrences or processes. This approach is founded on the idea that reviewing and combining previously known and successfully documented knowledge is the best way to solve a problem. Then, the phase will be directed, and the earlier stages will be fixed. Better final results will arise as a result. Table 3 compares the items mentioned above:

**Table 3. Comparing the modelling**

| No | Model type | The main Model | Specimens |
|---|---|---|---|
| 1 | Mathematical Model | A formulaic abstract of relevant elements, and then analyzing the network security | AHP model, Bayesian network, Fuzzy set/Rough set, Reliability/Survivability model |
| 2 | Random/stochastic Model | The investigation of behavioural aspects for security evaluation is based on an interactive description. | Petri net, Game theory, Markov model, Attack model, D-S evidence model, Risk diffusion model |
| 3 | Biologically-inspired Model | Multi-layer nonlinear fitting is used to analyze the security situation in conjunction with artificial intelligence. | Neural network, Artificial immunity, Genetic algorithm/Particle swarm optimization. |

Additionally, a technique was discussed for utilizing a convolutional neural network model (CNN) (in eight layers) in the Invalid source specified. The developers of this approach built their Model based on learning the prominent features of botnet networks and detecting malicious activity using energy consumption data.

**4) Results**

We shall define modelling and the mathematical Model initially at the commencement of the analysis:

***Mathematical Modelling***: Mathematical modelling applies mathematical ideas and language to describe a system. Mathematical models are commonly utilized by physicists, engineers, statisticians, operational research analysts, and economists in addition to the natural sciences (such as physics, biology, geosciences, and meteorology) and technical disciplines (such as computer science and artificial intelligence). A model can aid system explanation, component effects research, and behaviour prediction [14].

In a mathematical model, a number of variables are used to represent inputs, outputs and internal states, and a set of equations and inequalities to describe their interaction. In other words, this Model is a system description combining language and mathematical concepts. A model can aid system explanation, component impact research, and behaviour prediction.

The suggested strategy is shown in Figure 8 below:

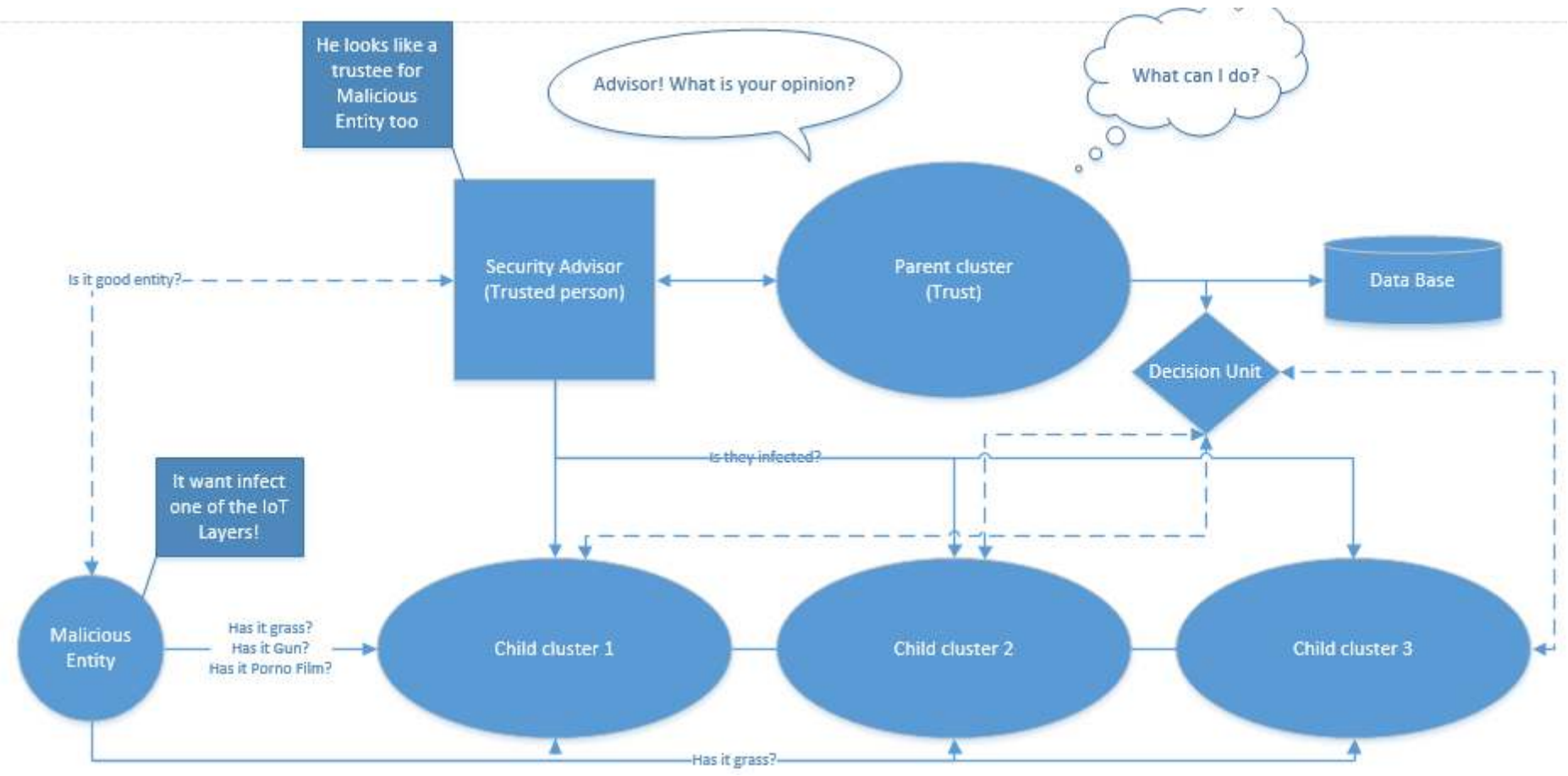

**Figure 8. The proposed plan**

The following components may be found in the diagram above, arranged from left to right:

- Malicious Entity
- Security Advisor
- Parent Cluster
- Decision Unit
- Data Base
- Child Cluster 1, 2, …

A map similar to Figure 9 may be created by contrasting the suggested strategy with the drawing mechanism in Figures 7 and 2:

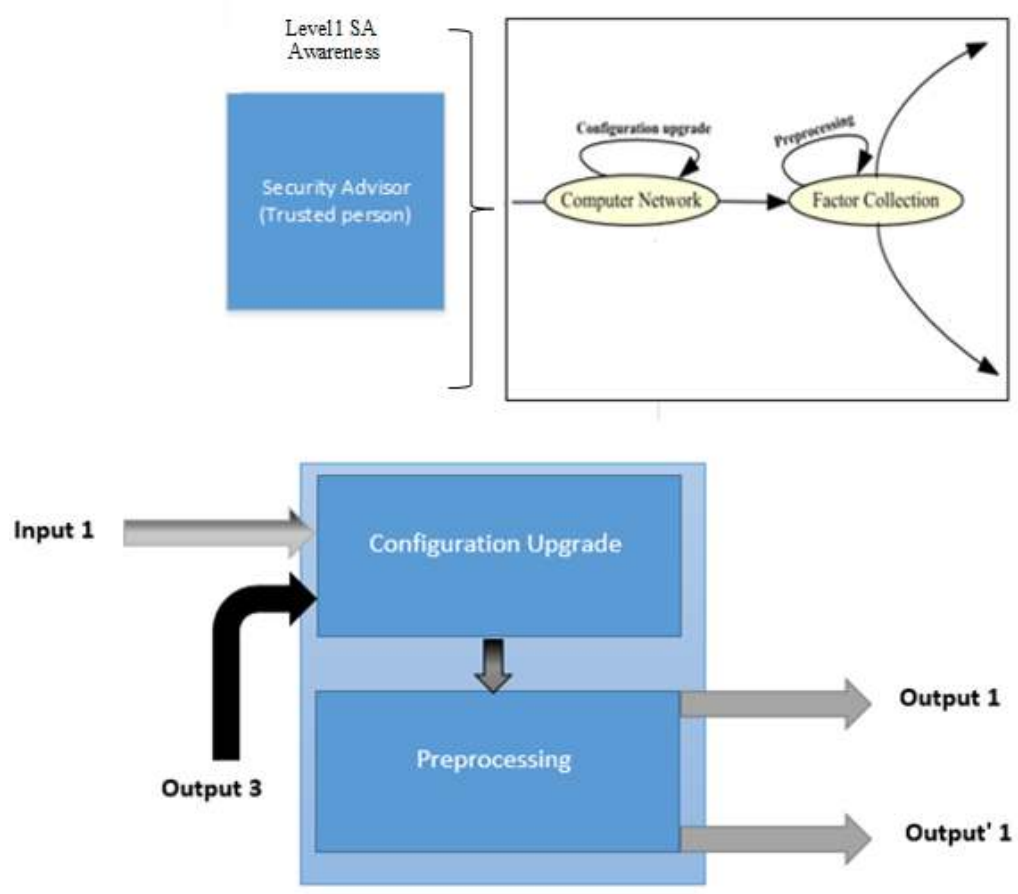

**Figure 9. Security advisor and its tasks**

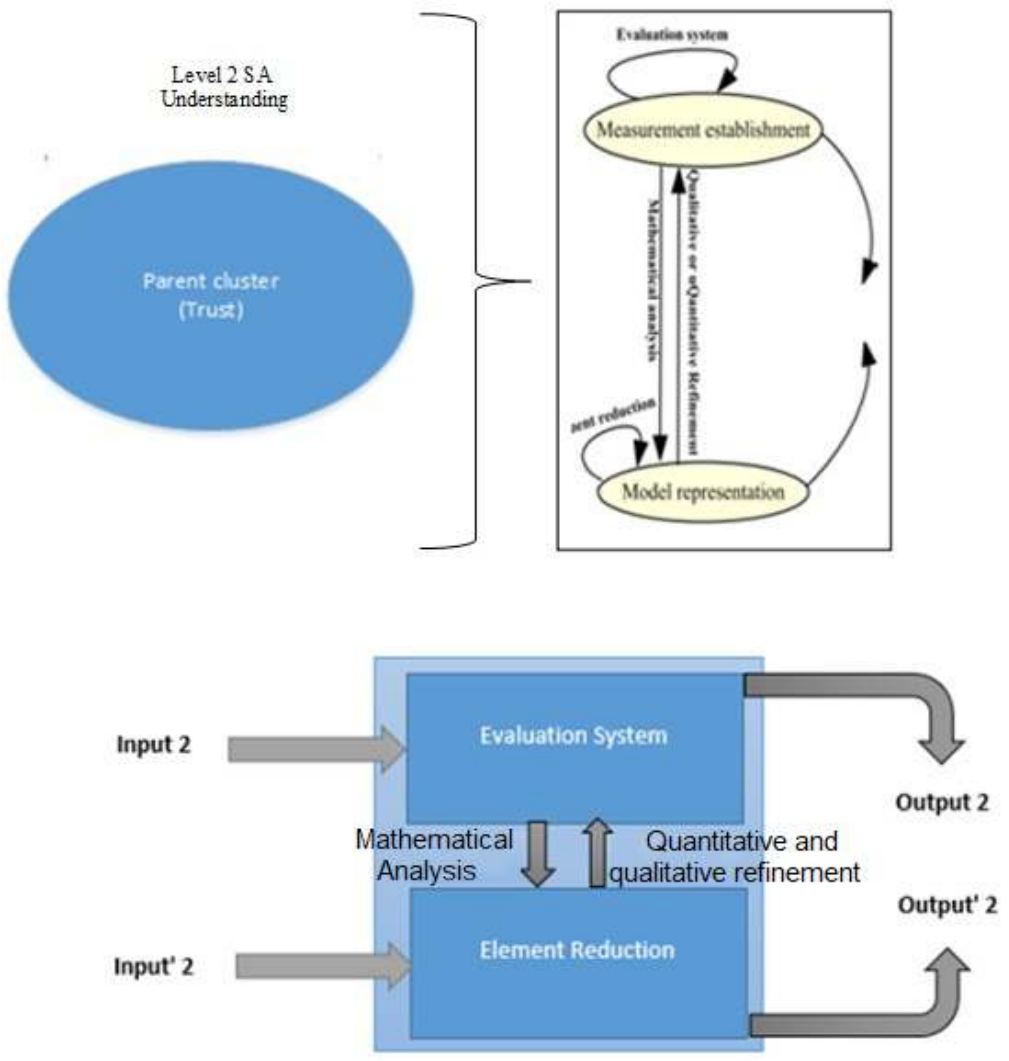

**Figure 10. The parent cluster and its tasks**

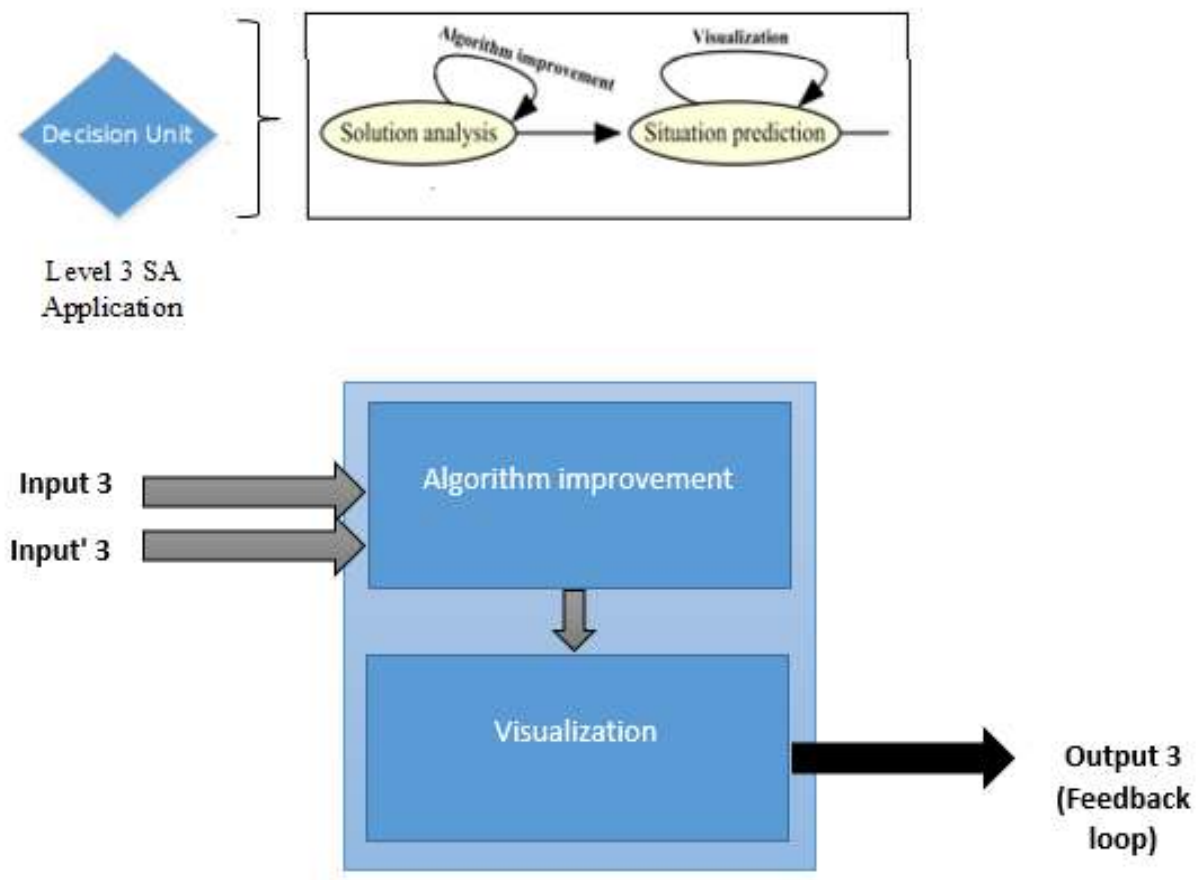

**Figure 11. The decision unit and its tasks**

In light of what was mentioned, we describe a three-layer model in [15] that can accomplish the objectives of trust management in the Internet of Things at each layer, as shown in Figures 2, 7, 8, and 12 (Fig.13):

- Layer 1: Awareness layer
- Layer 2: Perception layer
- Layer 3: Application and action layer

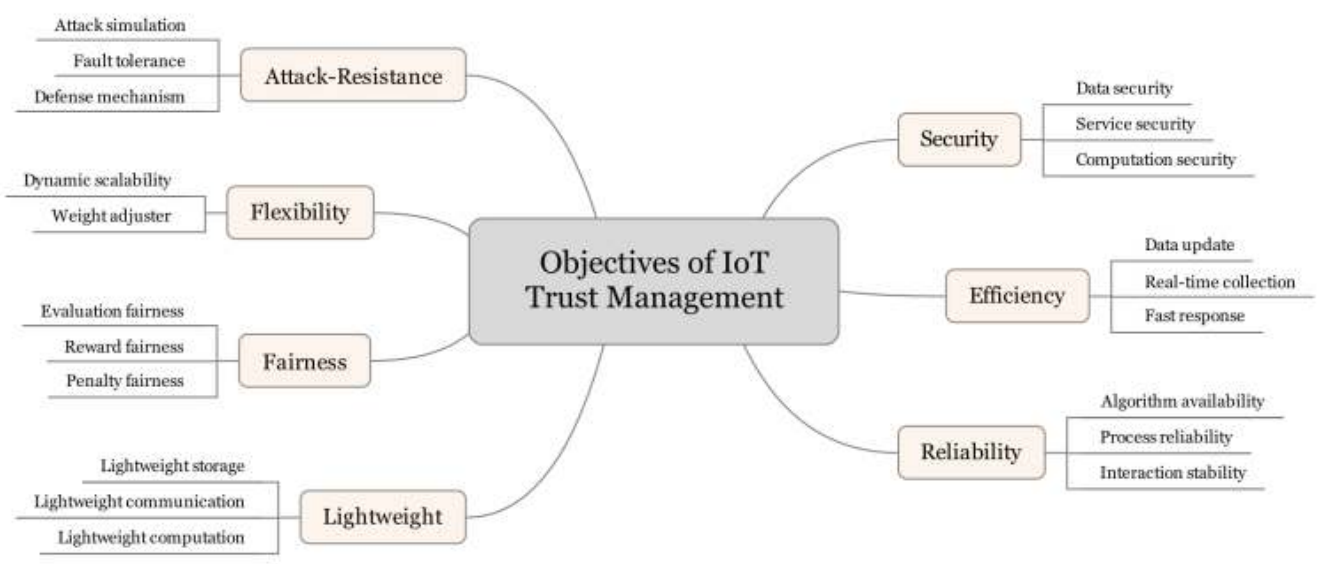

**Figure 12. Objectives of IoT trust management [5]**

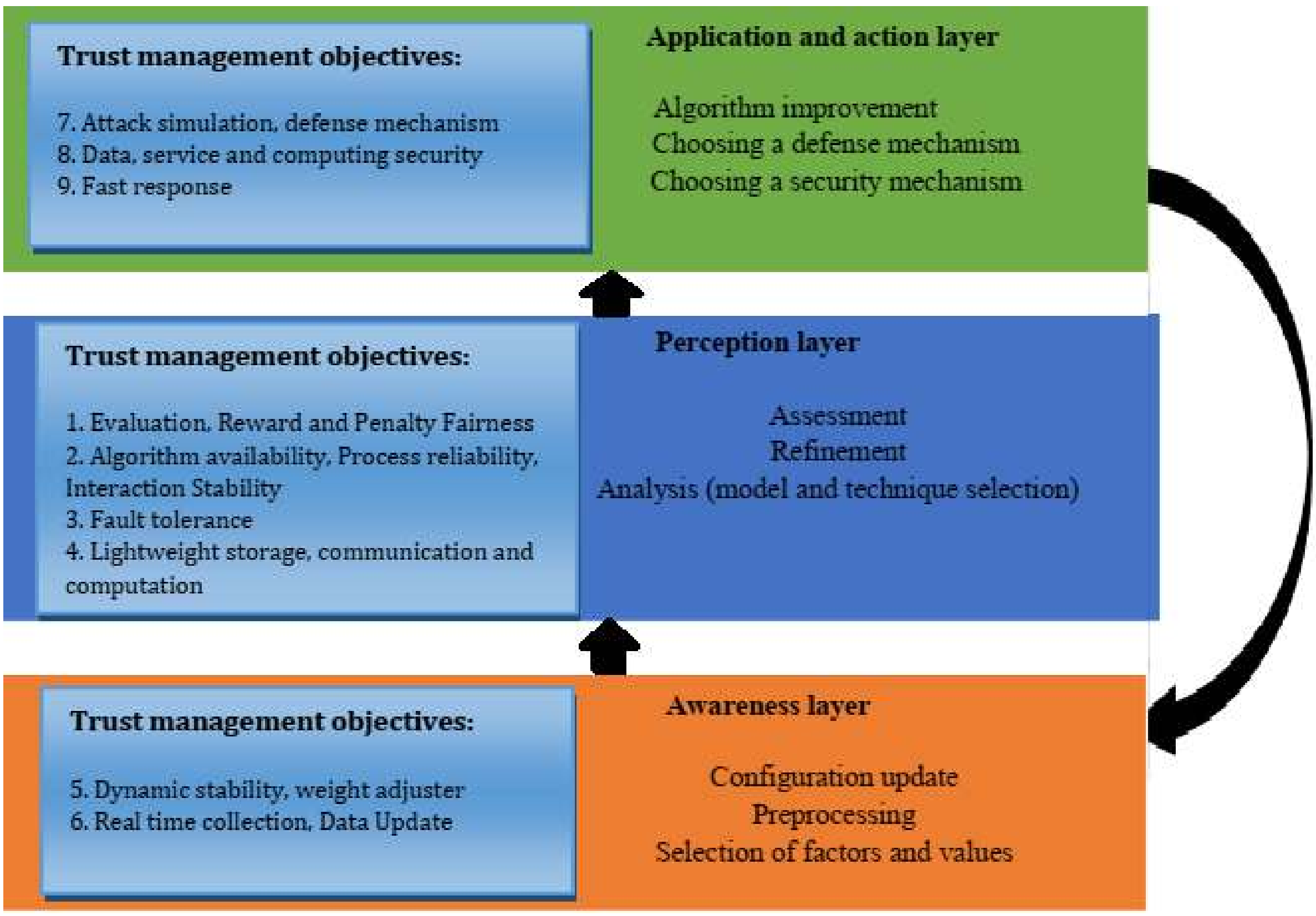

**Figure 13. Suggested three-layer**

A layer is described in [15] and in its suggested Model, as shown in Figure 14:

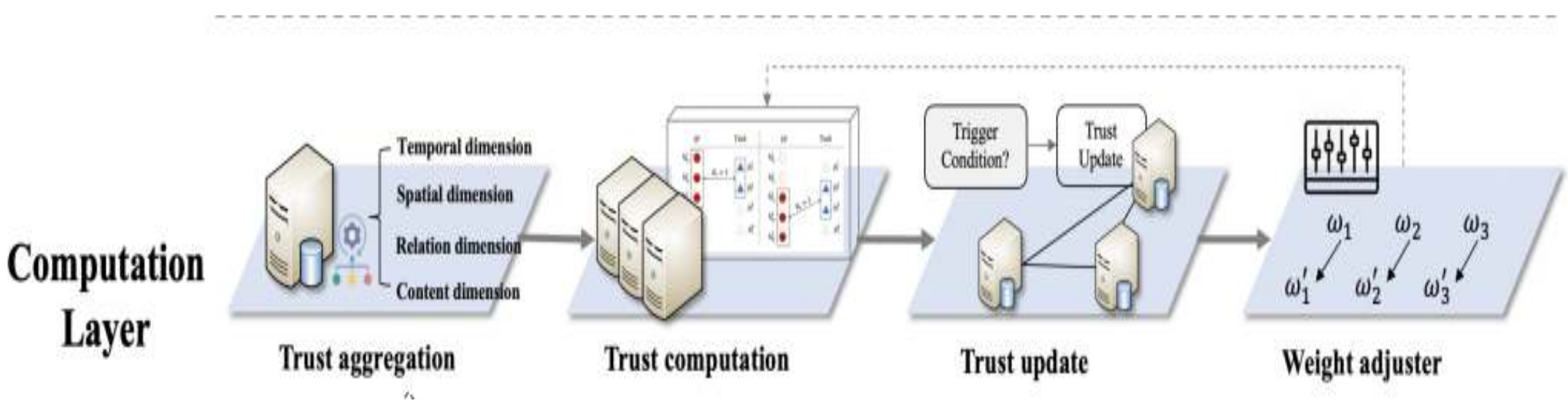

**Figure 14. Computation layer**

As shown, four steps are defined:

1. Trust aggregation

a. Temporal dimension
b. Spatial dimension
c. Relation dimension
d. Content dimension

2. Trust computation
3. Trust update
4. Weight adjuster

According to [15], there is a table identified as Table 4:

**Table 4. Confidence quantity**

| | | Trust indicator |
|---|---|---|
| Trust quantification | Computation method | Subjective logic |
| | | Bayesian inference |
| | | Fuzzy logic |
| | | Evidence theory |
| | | Analytic hierarchy process |
| | | Gray prediction |

## 6. Discussion

Given the presented Model, Table 5 is displayed as follows:

**Table 5. Techniques and models in each layer**

| Layer | The desired unit in the proposed Model | The utilized Model | Suggested technique |
|---|---|---|---|
| Awareness | Security Advisor | Evidence Theory<br>Subjective logic | Refer to Table 8 |
| Perception | Parent cluster | Fuzzy Logic<br>Gray prediction | Refer to Table 8 |
| Application and Action (using the CRIME model to identify a botnet) | Decision unit | Analytic hierarchy theory<br>Game theory | Refer to Table 8 |

The authors primarily defined the following three indicators in [16], expressed as the ratio of the optimized and original design (compared to their previous design) in a specific performance aspect: ratios of costs, times, and errors. However, as [16] concentrates on cars and drones, we take into account the following three ratios or rates for our proposed plan:

- Energy ratio with E parameter
- Time ratio with T parameter
- False ratio with F parameter

We complete Table 5 as Table 6 by using the three ratios mentioned above:

**Table 6. Important parameters**

| Layer | The desired unit in the proposed Model | The utilized Model | Important parameter | Suggested technique |
|---|---|---|---|---|
| Awareness | Security Advisor | Evidence Theory<br>Subjective logic | **Time** | Refer to Table 8 |
| Perception | Parent cluster | Fuzzy Logic<br>Gray prediction | False | Refer to Table 8 |
| Application and Action<br>(Using the CRIME model to identify a botnet) | Decision unit | Analytic hierarchy theory<br>Game theory | **Time and energy** | Refer to Table 8 |

The following techniques are evaluated and recommended for usage in the application and action layer under Table 7:

**Table 7. Comparing the trust management techniques in IoT**

| Technique | Achievements | Constraints |
|---|---|---|
| SMA | Because it extracts text and quantitative data from IoT devices | Since it employs textual and numerical data to find resources and determine the trust |
| | over the network, it is more reliable. | score, there is an increase in computational overhead. |
| ABAC | It enhances scalability, offers secure authorization, and expedites the decision-making process. | If a node interacts with several nodes at the same time, it is difficult to predict its reliability. |
| DCTEPF | This Model may be used to screen out irrelevant data. | It is not appropriate for managing background data for trust predictions. |
| ANTs | It monitors the network to identify malicious nodes. | It faces some problems in dividing the network into trusted areas. |
| CTM-IoT | It establishes reliable communication between all IoT devices. | Since it is not evaluated/compared with other plans, its superiority over existing techniques is ambiguous. |
| IoT-HiTrust | It achieves appropriate trust properties by considering attacks in an extensive IoT system. | It may be under the control of intruders since it does not include intrusion detection. |

The technique column is determined as follows:

**Table 8. Suggested technique for each layer**

| Layer | The intended unit in the suggested Model | Utilized Model | Type | Important parameter | The Suggested technique |
|---|---|---|---|---|---|
| Awareness | Security Advisor | Evidence Theory | **Random** | **Time** | Clever Middleware Architecture (SMA) |
| | | Subjective logic | **Probability logic** | | |
| Perception and decision | | Fuzzy Logic | **Mathematical Model** | | Attribute-based Access Control(ABAC) |
| | | | | | |
| | Parent cluster | Gray prediction | | **False** | DCTEPF (The data-centric trust evaluation and prediction framework) |
| Application and Action (Using the CRIME model to identify a botnet) | Decision unit | Analytic hierarchy theory | **Mathematical Model** | **Time and energy** | ANTs(Application-Driven Network Trust Zones) |
| | | Game theory | **Mathematical Model** | | |

- Subjective logic is appropriate for modelling and analyzing circumstances containing uncertainty and largely unreliable sources.
- Bayesian inference is a statistical inference technique that uses Bayes' theorem to update a hypothesis's probability when new information or evidence becomes available. Bayesian inference is a crucial statistical method, particularly in mathematical statistics (mathematical Model).
- Fuzzy logic is an approach to variable processing that allows multiple possible truth values to be processed through a variable.. To get accurate results, fuzzy logic uses a heuristic approach to handle issues with an open and imprecise range of inputs (mathematical Model).
- Evidence Theory has been applied to demonstrate expert system uncertainty, particularly in diagnosis. It applies to decision analysis (stochastic Model).
- Using mathematics and psychology, analytical hierarchy theory, often known as the Analytic Hierarchy Process (AHP), is a way to organize and evaluate complicated choices. By defining criteria and options and connecting those components to the overarching objective, AHP offers a logical framework for a decision need (mathematical Model).
- One of the most crucial time series prediction techniques, grey prediction, is utilized to resolve uncertainty issues with sparse data and insufficient information.

According to what was said, we can now present Figure 15 with more details in the figure below. We call this Model: the "Safe House Model or SHoM":

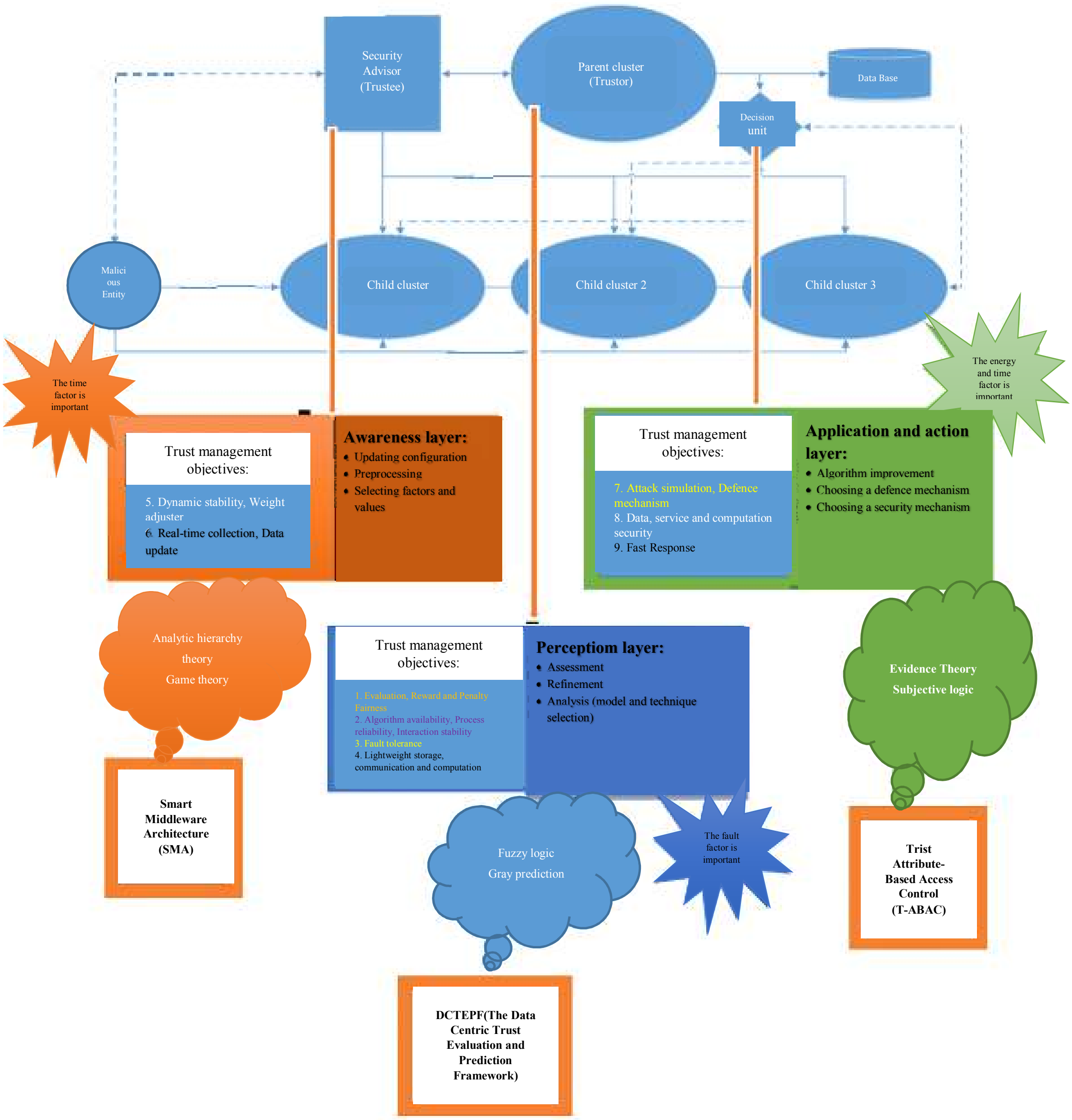

**Figure 15. "Safe House Model" or SHoM**

Table 9 describes the primary units of Figure 15

**Table 9. Explanation of the primary units of Figure 8**

| unit name | important factor | description | output |
|---|---|---|---|
| **Security advisor** | time | The awareness layer where "configuration update, preprocessing and | cannot be identified and is suspicious |
| | | selection of factors (values)" are performed.<br>In the proposed plan and in the Security Advisor unit, we use the SMA technique, in which the weighting system is used. As can be seen in the figure, the output of this unit is considered as the input of the Parent Cluster unit in the first processing. For trust evaluation, semantic discovery is the most important part of SMA, which uses text and numerical information provided by the communication device to perform semantic discovery. | |
| **Parent cluster** | Fault/error | This layer is chosen as the understanding layer where "evaluation, refinement and analysis (model and technique selection)" is performed in the Parent Cluster unit and according to the input from the Security Advisor unit. We consider the following parameters (T represents trust); here, the error factor is very important and influential. Here, the goal (g) is trust, and we are facing two types of trust: 1. Dependency Trust: Achieving goal g relies on the Security Advisor. 2- Realization trust: Security Advisor participation is necessary to perform the task (by the parent cluster). | To the advisor unit:<br><br>• Trusted, update yourself.<br>• Not trusted, wait for my reply and update yourself |
| | | | To the decision-making:<br><br>Unit. With the analysis done, the case is highly suspected to be a BOT malware attack. How should I react? |
| **Decision** | Time and Energy | This layer is selected as the application and action layer where "algorithm improvement, defence mechanism selection, security mechanism selection" are performed. This unit, according to the reliable and trusted output results from the "parent cluster", decides what reaction should be taken in front of the potential threat, which should be fast and optimal. | To the parent cluster unit (dependency trust):<br><br>Attack detected, activate 'X' defence mechanism. |
| | | | To child clusters (participation trust):<br><br>Defence mechanism 'X' has been activated by the parent cluster; take necessary security measures. |

## 7. Conclusion

The present study introduces an innovative hybrid mental model called SHoM, which addresses the critical challenge of botnet-driven DDoS attacks in the complex Internet of Things (IoT) environment with a multidimensional and integrated approach. SHoM fundamentally differentiates itself from conventional models in:

1. Trust Management Integration: Intelligently integrates trust management security concepts as a fundamental pillar, which is often overlooked in competing designs. This integration enables dynamic and context-aware assessment of the trustworthiness of nodes and data.
2. Self-organizing and perceptual structure: Having the ability to adapt and learn from the operational environment dynamically. This structure allows SHoM to adapt to evolving attack patterns and changing conditions of IoT networks without relying on a fixed configuration or centralized reference.
3. Multidimensional approach: Simultaneously addressing different dimensions of the DDoS threat (e.g., detection, sourcing, mitigation) and factors affecting IoT security (resource constraints, scalability, heterogeneity) in a single framework, as opposed to common one-dimensional or island-like approaches.

This unique combination (self-organisation, dynamic perception, and integrated trust management) has the potential to revolutionise IoT security. SHoM not only provides a more effective defence against botnet-based DDoS attacks but also opens up new research areas at the intersection of dynamic trust management, self-organizing systems, and resilient cybersecurity in IoT, suggesting novel perspectives for securing critical yet vulnerable IoT infrastructures. As a next step, the author plans to test and evaluate the SHoM model in more realistic and diverse IoT scenarios. These evaluations will focus on objectively measuring the benefits (such as detection accuracy, response speed, resource efficiency, scalability, resistance to sophisticated attacks) and identifying potential limitations or drawbacks compared to leading solutions. The results of these evaluations, along with more detailed analyses of the Model's performance and future improvements, will be presented to the IoT security community in subsequent research papers. The success of SHoM could be a cornerstone for a new generation of intelligent and resilient security systems in the IoT.